# The band calculation and local state analysis for armchair graphene-like nanoribbons with line defects

Yang Xie, Zhijian Hu, Wenhao Ding, Hang Xie[*]

College of Physics, Chongqing University

**Abstract**

In this paper we propose an analytical method to calculate the band structures of graphene-like nanoribbons of the armchair type with arbitrary line defects or uniaxial strains. The model is based on the tight-binding model and the standing wave assumption for the armchair nanoribbons. It gives accurate band results for large supercell systems. Within this method, we analyze different local states near the line defect in the graphene and boron-nitride nanoribbons. We also derive the analytical expression for the local states in a semi-infinite graphene nanoribbon by this method and the transfer matrix technique. The criteria condition for the local states in the semi-infinite nanoribbons is also given.

## 1. Introduction

Graphene, as a novel two-dimensional material, has attracted a lot of research interests in this decade since discovered in 2003[1]. It has many special electronic, optic and magnetic properties due to its energy band structure [2-3]. For example, graphene has the Dirac-like electron with the Klein tunneling effect [4-5]. And it has the half integer quantum Hall effect [6].

With the restriction in one dimension, the bulk graphene forms the graphene nanoribbons(GNR）. Armchair GNR (aGNR) and zigzag GNR (zGNR) are the two basic types [7]. There are edge states on edges of the zGNR [7].The early tight binding(TB) model shows that the edge state appears near the Fermi level, and where a large density of states exist [7,8]. Considering the Coulomb repulsion of the electrons with opposite spins by the Hubbard model, or using the first-principles calculation, it is found that the zGNR actually has a certain energy gap near the Fermi level and electrons with opposite spin are separately positioned on the two edges [9,10]. Some half-metal structures can be prepared by the modulated edges or doped atoms in the zGNR systems [11-13].

For aGNR, the TB calculation results show that they may be conductors and semiconductors [7,14]. The first-principles results show that for the metal-type aGNR in the TB model, in fact, there is still a very small gap [15,16]. However, the TB method is still a brief and useful model. It employs the $p_z$ orbital near the Fermi level to describe the electrical properties of GNR. The results basically agree with the first-principles calculations. In another example of GNR with a uniform stress applied on it, the TB model also gives the roughly same results as the first-principles calculation [17,18].

In the study of aGNR, an analytical solution for the energy band is given before [19]. The low-energy expansion and the edge deformation influence on the energy gap are studied. Among these studies, the transverse wave function of the ribbon is assumed to be the standing-wave form. Later on someone further used this assumption to calculate the electron states of semi-infinite aGNR [20]. The electronic structure of a uniform aGNR block with a finite length is also studied by the similar analytical method and the first-principles calculation [21,22]. It is found that in the semi-infinite aGNR or aGNR block the zigzag edges have localized states similar to that of edge states in the zGNR edges, which decay in the interior region. These local states are also verified from the surface Green's function calculation of aGNR [23].

However, these analytical studies are only the simple applications of the standing wave assumption in the transverse direction of aGNR. For the inhomogeneous aGNR systems, this analytical expression is not applicable.

In this paper we develop an analytical method with the standing-wave assumption to the inhomogeneous aGNR supercells. Within this method, we can do the energy band calculation to investigate the electronic structure of aGNR in the presence of various defects or strain conditions. And we analyze the possible local states in detail. We further extend this method to calculate the semi-infinite aGNR and the boron-nitride nanoribbons. We get the analytic solutions of local states in the semi-infinite aGNR and the asymmetrical local states in boron-nitride nanoribbon with a line defect.

This paper is mainly divided into the following parts: the second part is the introduction to the model and theory, the third part is about the calculation results, including the band structure and localized states in aGNR with line defects and uniaxial strains. The local states in the semi-infinite nanoribbon and the boron-nitride nanoribbon are also analyzed. The conclusion is in the last part.

## 2. Theory and models

### 2.1 Analytical analysis for the local states in the aGNR supercells

### 2.1.1 The wavefunction in aGNR with the form of standing wave

First we consider a simple aGNR as shown in Fig. 1(a). There are two types of carbon atoms (A and B) in one unit cell. In the tight binding model, the wavefunction can be written as the following form [19]

$$\psi = C_A \sum_A \sum_{j=1}^N e^{ikx_A} \phi_A(j) |A\rangle + C_B \sum_B \sum_{j=1}^N e^{ikx_B} \phi_B(j) |B\rangle. \tag{1}$$

Where the factors $e^{ikx_{A_j}}$ and $e^{ikx_{B_j}}$ are from the translational invariance in the horizontal direction. In the vertical direction, with the hard-wall boundary [19],

$$\phi_A(0) = \phi_B(0) = 0,$$

$$\phi_A(N+1) = \phi_B(N+1) = 0, \tag{2}$$

the following standing-wave-formed wavefunction is obtained

$$\phi_A(j) = \phi_B(j) = \sin\left(q_y j \frac{\sqrt{3}}{2} a\right). \tag{3}$$

With this form of wavefunction and the hard wall boundary condition, we have

$$q_y\left(N+1\right)\frac{\sqrt{3}}{2}a = p\pi \,,$$ (4)

where $p = 1, 2, ..., N$. Thus the wavefunction in y direction can be written as

$$\phi_A\left(j\right) = \phi_B\left(j\right) = \sin\left(\frac{p\pi j}{N+1}\right).$$ (5)

Then we employ this form of wavefunction into the Schrödinger equation with the tight-binding approximation and the eigenvalue equation is obtained for the band calculation [19].

This standing-wave assumption of the wavefunction can reduce any uniform (in y direction) aGNR system into a 1D system. Fig 1(b) shows such reduced 1D chain system. With this method, we can calculate for some large system with a much less computation load.

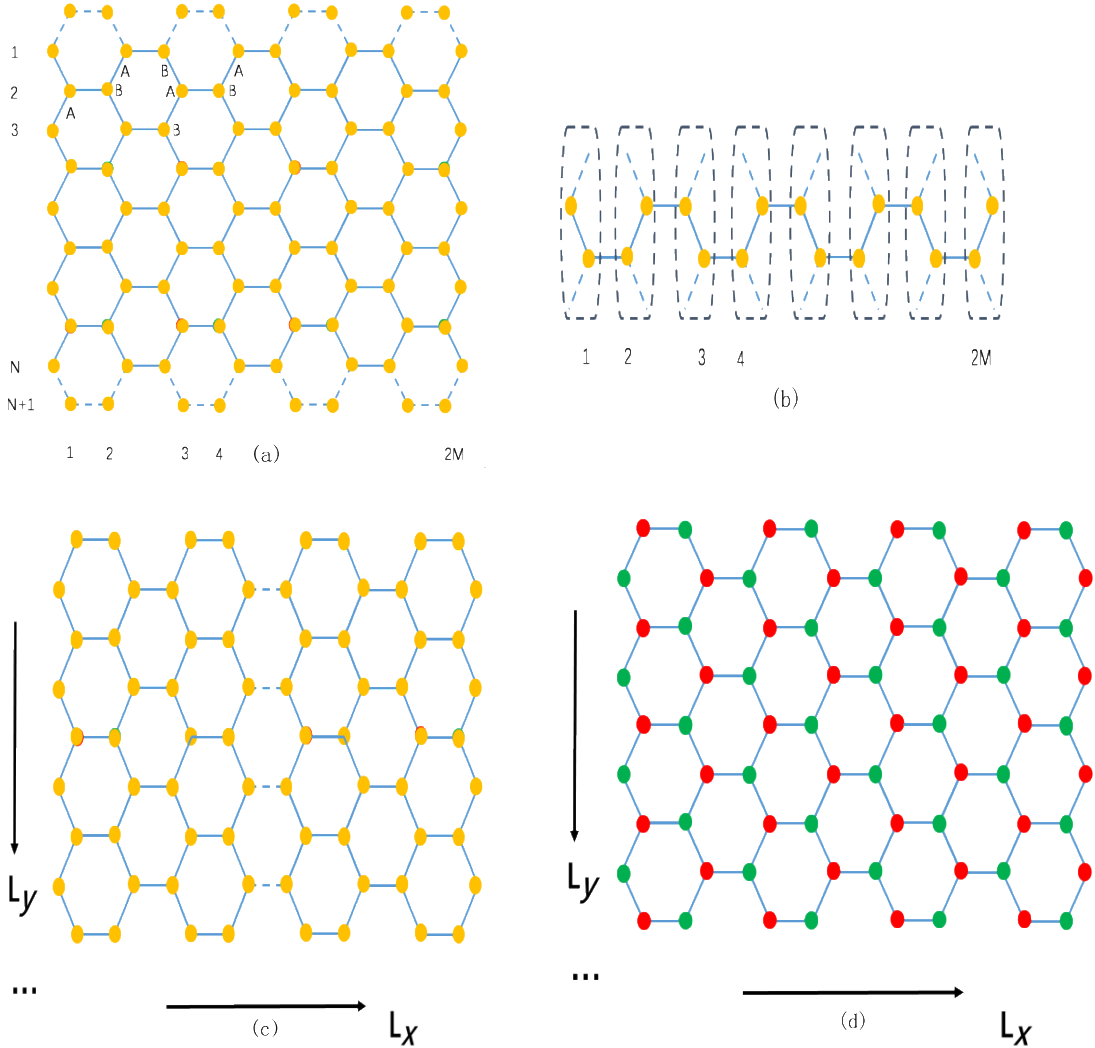

Fig. 1 The atomic structures for the armchair graphene or graphene-like nanoribbons. (a) The simple aGNR, each unit cell contains n numbers of A and B carbon atoms. The dashed lines in the top/bottom region indicate the boundaries of the standing wave. (b) The reduced 1D aGNR supercell. (c) The aGNR supercell with a line defect (indicated by the dashed lines in the middle region). (d) The supercell of a boron-nitride nanoribbon. The boron atoms are green and the nitride atoms are red.

### 2.1.2 The band calculation for aGNR supercells

Now we generalized this method to the supercell system, including the supercell with some line defects. This means the system is still uniform in the vertical direction, but the translation invariance in the horizontal direction is not satisfied. Furthermore, from Fig. 1(c) we see that the line defect destroys the six-member ring in one unit cell and the A-type atoms (or B-type atoms) in the original six-member ring are not equivalent. So we have to separate the original unit cell into two new sub-unit cells (Fig. 1(b)). Also due to the broken translational symmetry, in the new model the superposition coefficients of $A_n$ (and $B_n$) in different new sub-unit cells are not a constant. These coefficients are involved in a set of equations for eigenvalue calculation.

We assume that the supercell system has $N$ columns in y direction and $2M$ sub-unit cells in x direction. In the TB approximation, the wavefunction can be written as

$$|\psi\rangle = \sum_{n=1}^{2M} [\sum_{j=1}^{N} \phi_A(j) \cdot A_n |A\rangle + \sum_{j=1}^{N} \phi_B(j) \cdot B_n |B\rangle] , \tag{6}$$

where $\phi_A(j)$ and $\phi_B(j)$ are the coefficient components of A-type atom and B-type atom in the $j^{th}$ column of the sub-unit cell (see Eq. (5)). $A_n$ and $B_n$ denote the coefficient components of A-type atom and B-type atom in the horizontal direction. Under the TB approximation, the Hamiltonian is written as

$$H = \sum_i \varepsilon_i |i\rangle \langle i| + \sum_{<i,j>} t_{i,j} |i\rangle \langle j| , \tag{7}$$

where $<i,j>$ denotes the nearest neighbors. If we choose $\varepsilon_i = 0$ and $t_{i,j} = t$, and put Eq. (6) and (7) into the Schrödinger equation, we have the following equation for the A-type atoms in an aGNR supercell,

$$E\phi_A(j) A_n - t\phi_B(j) B_{n+1} - t\phi_B(j+1) B_n - t\phi_B(j-1) B_n = 0 . \tag{8}$$

Similarly, we have the following equation for the B-type atoms in an aGNR supercell

$$E\phi_B(j) B_n - t\phi_A(j) A_{n-1} - t\phi_B(j+1) A_n - t\phi_B(j-1) A_n = 0 . \tag{9}$$

Substituting Eq. (5) into Eqs. (8) and (9), and with some derivation, we have

$$EA_n - tB_{n+1} - 2t\cos\left(\frac{p\pi}{N+1}\right) B_n = 0 , \tag{10a}$$

$$EB_n - tA_{n-1} - 2t\cos\left(\frac{p\pi}{N+1}\right)A_n = 0 . \tag{10b}$$

The two equations above are $j$ independent. Then we employ the Bloch boundary condition

$$B_{2M+1} = e^{ik2aM}B_1 , \tag{11a}$$

$$A_0 = e^{-ik2aM}A_{2M} . \tag{11b}$$

Substituting this Bloch boundary condition into Eq. (10), we obtain the eigenvalue matrix equation which involves the Bloch wave vector $k$. For an aGNR supercell with 2M unit cells in the x direction, there are 4M unknowns ($A_n$ and $B_n$). So we have a 4M*4M matrix equation with a number of 4M eigenvalues. For each matrix equation, we find that N/2 number of $p$ values (Eq. (5)) can be chosen, which corresponds to the number of the standing-wave modes in the y direction (we will discuss more details about this later). So the total number of eigenvalues is 2M*N, which exactly equals to the number of energy bands in the common TB method.

The equations above is for the uniform (in x direction) aGNR supercell. For the supercell with a line defect, we assume that the defect only change the hopping integral $t$ [26,27,29], Eqs. 10 (a) and 10 (b) are modified as

$$EA_n - t_n^{A,1}B_{n+1} - 2t_n^{A,2}\cos\left(\frac{p\pi}{N+1}\right)B_n = 0 , \tag{12a}$$

$$EB_n - t_n^{B,1}A_{n-1} - 2t_n^{B,2}\cos\left(\frac{p\pi}{N+1}\right)A_n = 0 , \tag{12b}$$

where $t_n^{A,1}$ and $t_n^{A,2}$ correspond to the hopping integrals from the A-type atoms. $t_n^{A,1}$ is the hopping integral from A-type atom to the horizontal forward B-type atom; $t_n^{A,2}$ is the hopping integral from A-type atom to the top (bottom)-left B-type atoms. $t_n^{B,1}$ and $t_n^{B,2}$ are defined similarly(see Fig. 1b) . We assume the line defect only changes the hopping integrals in the horizontal direction. It is easy to see that $t_{n-1}^{A,1} = t_n^{B,1}$ and $t_n^{A,2} = t_n^{B,2}$ .

With the approach above, we set up a new one-dimensional model for the band structure calculation of an aGNR supercell system.

### 2.2 Local states in a semi-infinite aGNR

We consider a simple semi-infinite aGNR as shown in Fig. 1(a). According to the theory we generalized before, Eqs.10 (a) and 10(b) also hold except the equation the boundary sub-unit, which is modified as

$$EB_1 - 2t\cos(\frac{p\pi}{N+1})A_1 = 0 . \tag{13}$$

From Eqs. 10 (a) and 10 (b), we obtain a transfer matrix equation

$$\begin{pmatrix} B_{n+1} \\ A_{n+1} \end{pmatrix} = \mathbf{T} \begin{pmatrix} B_n \\ A_n \end{pmatrix} = \begin{pmatrix} -\tau t^{-1} & Et^{-1} \\ -Et^{-1} & t^{-1}E^2\tau^{-1} - t\tau^{-1} \end{pmatrix} \begin{pmatrix} B_n \\ A_n \end{pmatrix},$$  (14)

where $\tau = 2t\cos\left(\dfrac{p\pi}{N+1}\right)$. To solve the semi-infinite problem with this transfer matrix method,

we calculate the eigenvalues $\lambda_1$, $\lambda_2$ and the corresponding eigenvectors $[U_{11}, U_{21}]^{\mathrm{T}}$, $[U_{12}, U_{22}]^{\mathrm{T}}$ of $\mathbf{T}$. The two eigenvalues satisfy the condition $\lambda_1\lambda_2 = 1$. To get a solution of local (decaying) state, the eigenvalues must be real. We assume $\lambda_1 < 1$ and $\lambda_2 > 1$. Then we have

$$\begin{pmatrix} B_{n+1} \\ A_{n+1} \end{pmatrix} = \begin{pmatrix} U_{11} & U_{12} \\ U_{21} & U_{22} \end{pmatrix}^{-1} \begin{pmatrix} \lambda_1 & 0 \\ 0 & \lambda_2 \end{pmatrix} \begin{pmatrix} U_{11} & U_{12} \\ U_{21} & U_{22} \end{pmatrix} \begin{pmatrix} B_n \\ A_n \end{pmatrix}.$$

Iterating the formula above, we obtain the following result

$$\begin{pmatrix} B_{n+1} \\ A_{n+1} \end{pmatrix} = \begin{pmatrix} U_{11} & U_{12} \\ U_{21} & U_{22} \end{pmatrix}^{-1} \begin{pmatrix} \lambda_1^n & 0 \\ 0 & \lambda_2^n \end{pmatrix} \begin{pmatrix} U_{11} & U_{12} \\ U_{21} & U_{22} \end{pmatrix} \begin{pmatrix} B_1 \\ A_1 \end{pmatrix}.$$  (15)

Since we assume $\lambda_2 > 1$, to avoid the divergence of the wavefunction, the following condition is required

$$U_{21}B_1 + U_{22}A_1 = 0.$$  (16)

Combining the boundary equation (Eq. (13)) and Eq.(16) we have

$$\begin{pmatrix} U_{21} & U_{22} \\ E & -2t\cos\left(\dfrac{p\pi}{N+1}\right) \end{pmatrix} \begin{pmatrix} B_1 \\ A_1 \end{pmatrix} = \mathbf{D} \begin{pmatrix} B_1 \\ A_1 \end{pmatrix} = 0.$$

We see that $\det\mathbf{D} = 0$ is a sufficient condition for the existence of a local state. Our calculation indicates that such a condition can be satisfied only with $E = 0$. Such a conclusion is consistent with the work of Jiang and others [20]. When $E = 0$, we can easily obtain the local state solution from Eqs. (13) and (14).

$$A_n = 0,$$  (17a)

$$B_n = \left(-2\cos\left(\frac{p\pi}{N+1}\right)\right)^{n-1} B_1.$$  (17b)

For a convergent solution, it is apparent that $p$ satisfies the condition

$$2\cos\left(\frac{p\pi}{N+1}\right) < 1,$$

or

$$p > \frac{N+1}{3}.$$ 

(18)

We find this local state also exist in finite-sized graphene nanoribbons (or graphene quantum dot) [24]. In the literature [24], $\frac{2p\pi}{(N+1)a_0} - \frac{2\pi}{3a_0} > \frac{1}{L_y}$ is the sufficient condition for the existence of a local state for finite-sized graphene nanoribbons. As for a semi-infinite aGNR $L_y \to \infty$, we have $p > \frac{N+1}{3}$. This conclusion is consistent with Eq. (18). It is easy to understand that for the energy band beside the Dirac point, $E \propto \sqrt{k_x^2 + k_y^2}$. If $k_x$ is a pure imaginary number, $k_y$ must have a larger absolute value than that of $k_x$.

Finally, if A-type atom and B-type atom are different as shown in Fig. 1(d) in an armchair boron-nitride nanoribbon [25], Eqs.10(a), 10(b) and (13) are modified as

$$(E - \varepsilon_B)B_1 - 2t\cos(\frac{p\pi}{N+1})A_1 = 0,$$ 

(19a)

$$(E - \varepsilon_A)A_n - 2t\cos(\frac{p\pi}{N+1})B_n - tB_{n+1} = 0,$$ 

(19b)

$$(E - \varepsilon_B)B_{n+1} - 2t\cos(\frac{p\pi}{N+1})A_{n+1} - tA_n = 0,$$ 

(19c)

where $\varepsilon_A$ and $\varepsilon_B$ are the on-site energies of A-type atom and B-type atom. When $E = \varepsilon_B$, we can obtain a similar local state.

## 3. Results and discussions

### 3.1 Band calculation for simple aGNRs and aGNRs with uniaxial strain

Firstly we use Eq. (10) to calculate a simple aGNR for a benchmark. It is a supercell with M=2. The hopping integral of the aGNR is chosen as $t$=-2.7 eV, which agrees well with the results from the first principle calculations [12]. For the width with N=6 and N=8, we give the calculation results as below. They agree well with the band structures in the literature [19].

Our method can also calculate the band in aGNR with uniaxial strain [18]. As the method developed in the papers [17,18], the three bond vectors in the aGNR are changed with the uniaxial strain $\sigma$ in x-direction as $r_{ix} \to (1+\sigma)r_{ix}$; $r_{iy} \to (1-\nu\sigma)r_{iy}$, where $\nu = 0.165$ is the Poisson ratio [18]. The hopping integrals are scaled by the factor $\xi = (\frac{r_0}{r})^2$, where $r_0$ is the unstrained bond length and $r$ is the bond length with strain. With some derivation, the new hopping integrals in Eq. (12) are obtained as $t_n^{A,1} = t_n^{B,1} = t_0(1 - 2\sigma + 3\sigma^2)$ and

$$t_n^{A,2} = t_n^{B,2} = \left[\frac{1}{2}(1+\sigma^2)^2 + \frac{3}{2}(1-\nu\sigma)^2\right]^{-1} \approx t_0\left(1 - \frac{1}{2}\sigma + \frac{3}{2}\nu\sigma\right).$$ We choice the supercell

sizes with M=2 and N=23,24,25, which correspond to N+1=3q,3q+1,and 3q+2 cases[15]. We calculate the band gap in aGNR with different strain as shown in Fig. 2(c). We find there are periodic changes of the band gap with different strain. Our results agree well with the results from the perturbation theory calculations [18]. And our results come back to the unstrained cases [15] when $\sigma = 0$.

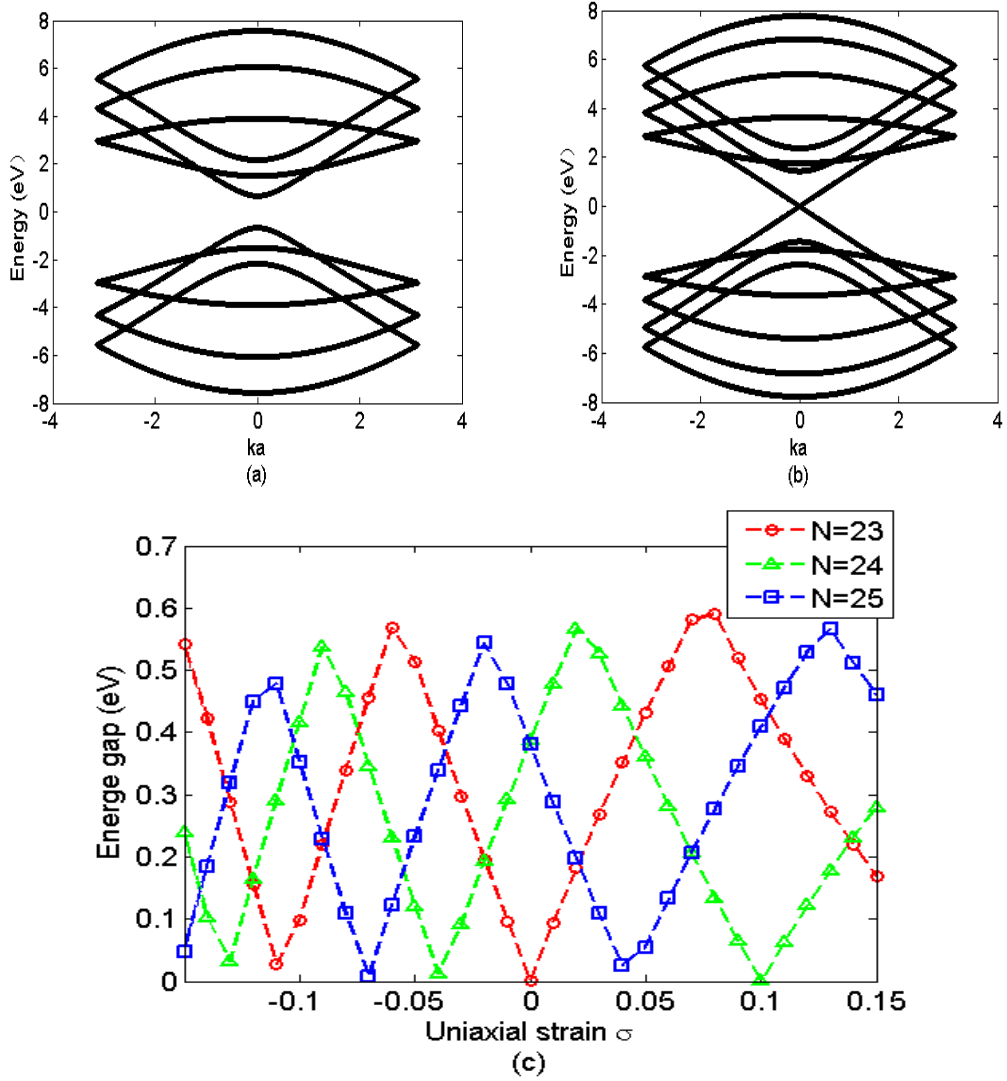

Fig. 2 The band structures of simple aGNRs (2M=2) and aGNRs with uniaxial strain in x-direction by the 1D-supercell method. (a) A simple aGNR with N=6; (b) A simple aGNR with N=8. a is the period of the supercell and a=3a$_0$, where a$_0$ is the lattice constant of aGNR; (c) the band gaps of aGNRs with uniaxial strain $\sigma$ in x-direction.

## 3.2 Band structures of aGNR with a line defect

With this supercell method, we now calculate the band structures of the aGNR supercell with a line defect as shown in Fig. 1(c). We choose the hopping integrals in the defect positions as $t_{n-1}^{A,1} = t_n^{B,1}$ =-0.5 eV. Other hopping integrals are set as the value of -2.7 eV. With Eq. (12) and the boundary condition (Eq. (11)), we set up a matrix equation for the eigenvalue calculation.

The calculated band structures are shown in Fig.3 below.

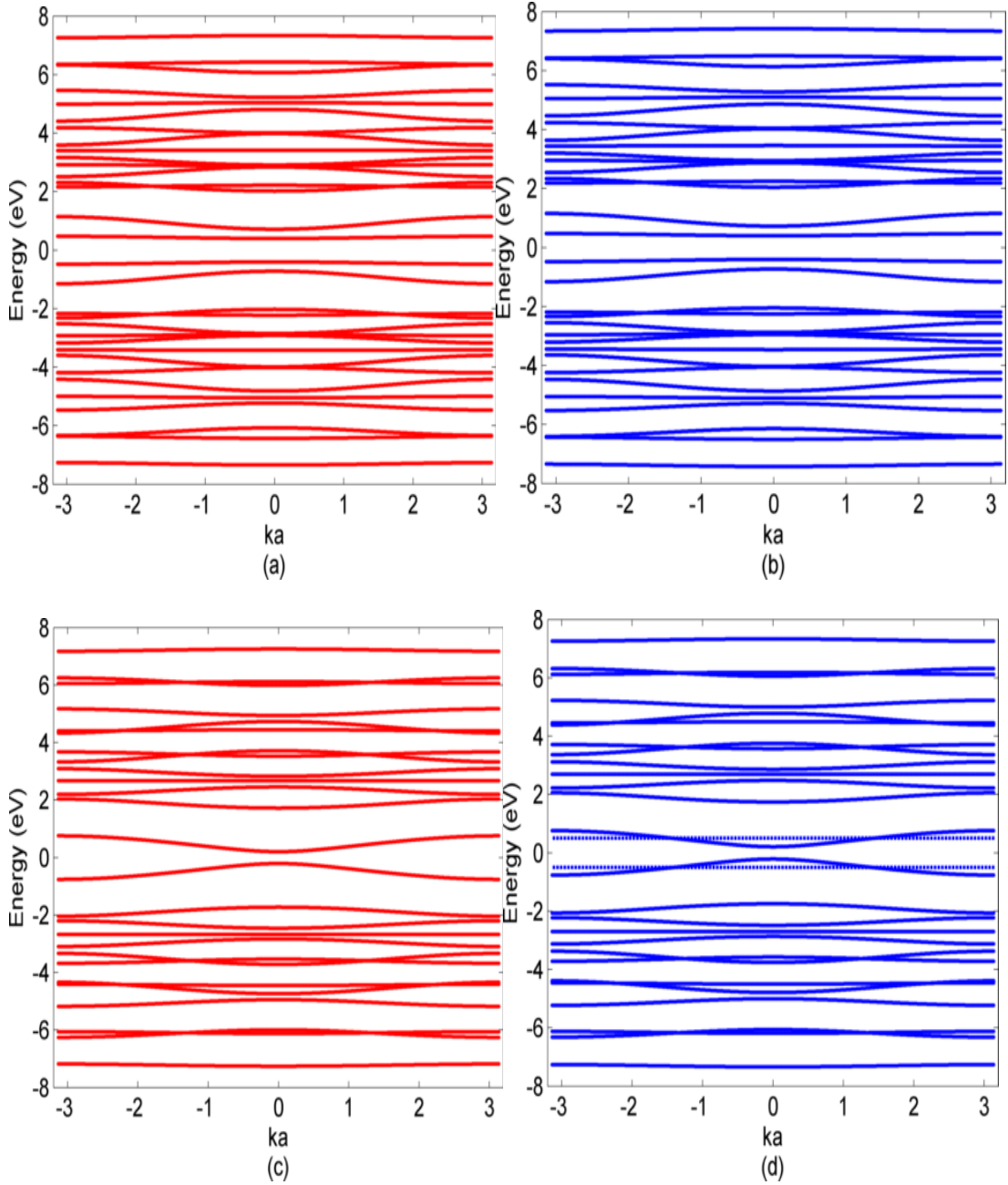

Fig. 3 The band results from the common 2D-TB method ((a), (c), red) and our 1D supercell method ((b), (d), blue)for aGNR supercell with a line defect. (a) and (b): The supercell size parameters are N=8, 2M=8; (c) and (d): The supercell size parameters are N=7, 2M=8. In (d) the dashed lines denote the bands which are need to be deduced artificially.

Fig. 3(a) and (b) are for the supercell with the size N=8, 2M=8. As stated before, we choose a number of N/2 $p$ values for the eigenvalue calculations. In each eigenvalue matrix, there are 4M eigenvalues. So the total number of the bands for this system is 64 (2M*N), which is the total number of the atoms in the aGNR supercell.

Now we study for another case: the supercell with the size N=7, 2M=8. In this case N is an odd number. We find if we chose (N+1)/2 or (N-1)/2 $p$ values, the total number of bands is 2(N+1)*M or 2(N-1)*M, which is not equal to the number of bands in the 2D-TB model (2M*N).

Here we give a detailed analysis for this discrepancy. As we stated previously, each $p$ value corresponds to a type of standing wave in the y direction in the 1D model. In a simple aGNR, A-type and B-type atoms are equivalent, so $p$ values can be chosen from 1 to N [19]. Here in our supercell (with line defects) case, A-type and B-type atoms are not equivalent. There are (N-1)/2 (or (N+1)/2) A-type atoms and (N+1)/2 (or (N-1)/2) B-type atoms in each separated unit cell. So they have different number of the corresponding standing waves. In this method, we find for each 4M*4M matrix equation, we have to choose (N+1)/2 $p$ values, thus we obtain a total number of 2M*(N+1) bands in the calculation. Then we have to artificially take out 2M bands for A-type atoms with $p$=(N+1)/2. The final number of the bands is 2M*N, which coincides with the results of the 2D-TB model.

In Fig. 3(c) we see that there are 7*8=56 bands from the 2D-TB calculation. From the 1D-supercell calculation, the total number of bands with (N+1)/2 $p$ values is 64, as shown in the blue lines in Fig. 3(d). Then we deduct 2*M=8 bands (dashed lines) for $p$=(N+1)/2 (each line is 4-fold degenerate), and the left bands are exactly the same as those in Fig. 3(c).

We notice that this 1D-supercell method can give the same band result as the 2D-TB method, only if the aGNR supercell is uniform in y direction. For a very large system our 1D-supercell method has a much small computation load, since it only needs to solve 4M*4M eigenvalue matrices, instead of solving a (2M*N)* (2M*N) large eigenvalue matrix.

### 3.3 The local states in the aGNR supercells

Now we begin to analysis the local states in these aGNRs. The edge states are well known in the zigzag GNR. Here we find that in the armchair GNR, if there is some line defect, there also exist such edge states, which are localized near the line defect. These local states result from the zigzag edge structures, which were discussed in the literatures before [22,29].

We use an aGNR supercell (N=23, 2M=8) with a line defect on the supercell boundaries. The hopping integrals in the defect are set as $t_1 = t_{n-1}^{A,1} = t_n^{B,1}$ =-0.5 eV. Fig. 4(a) draws the band structure near the Fermi level. The numbers in the band figure denote the $p$ values. We also use the 2D-TB method to draw the electron density distribution for the band which is most close to the Fermi level ($p$=9). Fig. 4(b) and (c) show the results with $k$=0. We also calculate for $p$=9 band with the $k$ value averaged in the first Brillouin zone. The electron distribution is very similar with the $k$=0 case (Fig. 4(b)). We see that the electron is localized near the line defect in the supercell. In y direction, there are 3 peaks for this edge state. We believe the STM experiment can also detect these edge states as well.

Then we tune the hopping integral $t_1$ to a larger absolute value (-1.0eV), the electron density distribution is shown in Fig. 4(d) and (e). We see when $|t_1|$ increased, the edge state become more delocalized and the peaks in y direction disappear. In the limit of $t_1$ equal to $t_0$ (-2.7 eV), the edge state will be replaced by a uniform state in x direction.

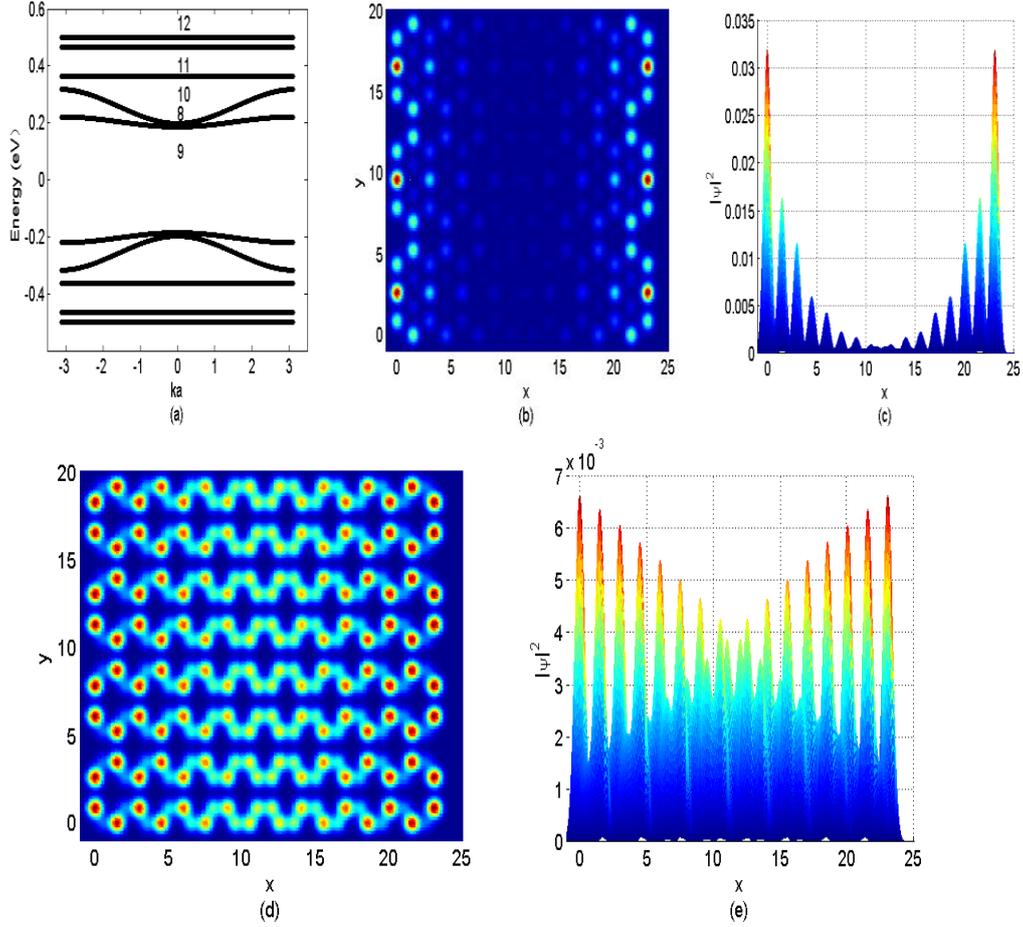

Fig. 4 The band structure and the electron density distributions for the edge state in a GNR supercell (N=23, 2M=8) with line defect. (a) The bands near the Fermi level, the numbers indicate the $p$ values in the 1D-supercell calculation; (b) and (c): The electron density distribution of the edge state for $p$=9 band with $k$=0 and $t_1$=-0.5 eV. (b) is the overhead view and (c) is the side view for the 3D plot. (d) and (e): The electron density distribution of the edge state for $p$=9 band with $k$=0 and $t_1$=-1.0 eV. (d) is the overhead view and (e) is the side view of the 3D plot.

From Sec. 2.2, we see that in a semi-infinite aGNR the localized state only exist when $p > \dfrac{N+1}{3}$. We find this condition also holds for the possible local states in the defect-involved aGNR supercells, in which the $p$ value is chosen close to the Fermi level, and the hopping integral $t_1$ is set to be about zero, like an isolate system. For example, we consider a supercell with the size N=17，M=18. In the 1D-supercell method, we calculate and plot the wavefunction (real part) in the x direction (Fig. 5(a) and (b), corresponding to the side view of the 3D electron density distribution). The defects position at the left and right boundaries of the supercell. We see that for the local edge states, $p$ value should satisfy $\frac{N+1}{3} < p \le \frac{N+1}{2}$. The upper bound come from the standing wave condition as stated before. In this case there are three available $p$ values ($p$=7,8,9). In Fig. 5 (a) and (b) we see that the wavefunction is localized near the two boundaries (line defect). From Eq. (17b) we know that a large $p$ value corresponds to a stronger decaying mode. In Fig. 5 we see the wavefunction with $p$=8 has a more localized behavior than that with $p$=7 as expected. We also check that when $p \le 6$, there are no local states in the supercell system.

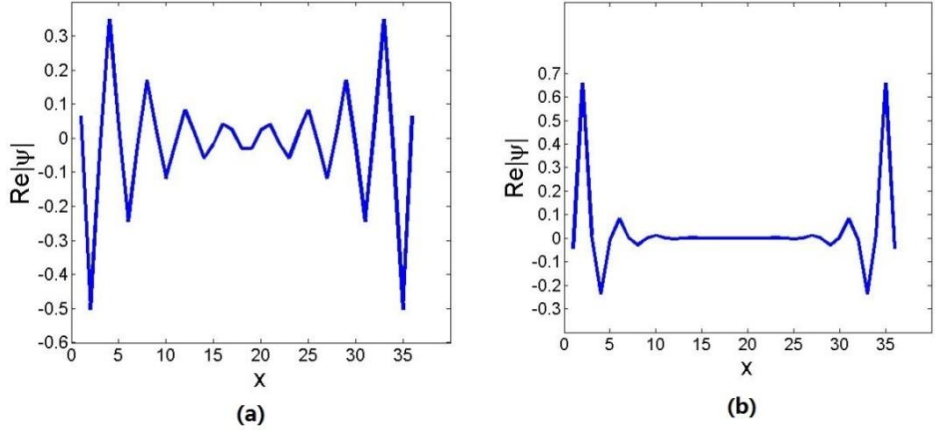

Fig.5 The local states in a long aGNR supercell with the isolate line defect ($t_1$=0). The size of the supercell is set as N=17, M=18. (a) The wavefunction of the 1D-supercell model with $p$=7; (b) The wavefunction of the 1D-supercell model with $p$=8.

### 3.3 Local states in the semi-infinite aGNR

Now we consider the localized state near the edge of a semi-infinite aGNR. As discussed in Sec. 2.2, the local state exist only at $E$=0 and $A_n$=0. We assume that $B_n$=1 and utilize the iteration relation in Eq. (17b) to calculate $B_n$ ($n>1$) for these local states in a semi-infinite aGNR. Then we use Eq. (5) to recover the 1D result into the 2D case. Fig. 6 gives an example for the local state in a semi-infinite aGNR with the width N=23 and the standing-wave number $p$=9.

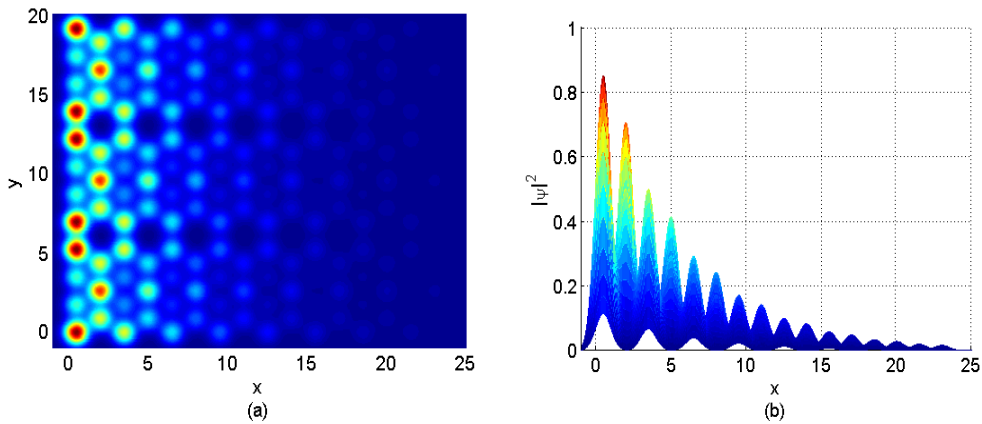

Fig. 6 The electron density distribution of the local state in a semi-infinite aGNR (N=23, p=9).

(a) is the overhead view and (b) is the side view of the 3D plot.

From Fig. 6 we see that this state is very similar to the local state in the aGNR supercell with a strong line defect. This is reasonable since a very long supercell with a strong line defect ($t_1=0$) can be regarded as two combined semi-infinite systems.

### 3.4 Local states in the armchair boron-nitride nanoribbon

Boron-nitride (BN) nanoribbon is very similar to the graphene nanoribbon, except that in the boron-nitride nanoribbon, A-type and B-type atoms are different (Fig. 1(d)). We can also use the method we introduced in Sec. 2 to calculate the energy band and local state. The parameters of the TB calculation are listed inTable1 [25].

| TB parameter | Value（eV） |
|---|---|
| $\varepsilon_N$ | -1.45 |
| $\varepsilon_B$ | 3.2 |
| $t_{B-N}$ | -2.45 |

Table1The Tight binding parameters for the boron-nitride nanoribbon

We firstly calculate the energy bands of two single-column armchair BN nanoribbons with the geometric sizes: (2M=2,N=6) and (2M=2,N=8), as shown in Fig. 7(a) and (b). Because of the symmetry breaking of A-type atom and B-type atom, a large band gap is open and the Dirac point is broken when N=8, compared with the band of aGNR.

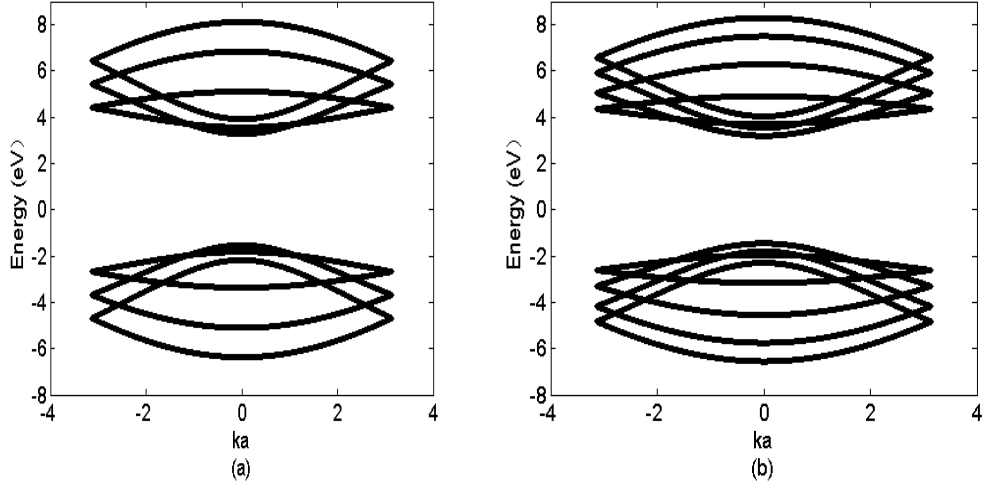

Fig. 7 The band structures of simple armchair BN nanoribbon (2M=2) by the 1D-supercell method. (a) N=6; (b) N=8. a is the period of the supercell. In this case, a=3a$_0$, a$_0$ is the lattice constant of armchair boron-nitride nanoribbon

Then we choose an armchair BN nanoribbon supercell with a line defect and the size parameter M=16, N=23 (see Fig. 1(d)). The hopping integrals in the defect positions are set as $t_1 = t_{n-1}^{A,1} = t_n^{B,1} = -1.0\,\mathrm{eV}$. We use our 1D-supercell method to calculate two local states for the bands near the Fermi level with $p$=8 and 9 (with $k$=0 in the reciprocal space), as shown in Fig. 8. This 1D-supercell method gives the same results as those in the 2D-TB method. The local state is asymmetry because the A-type and B-type atoms are different.

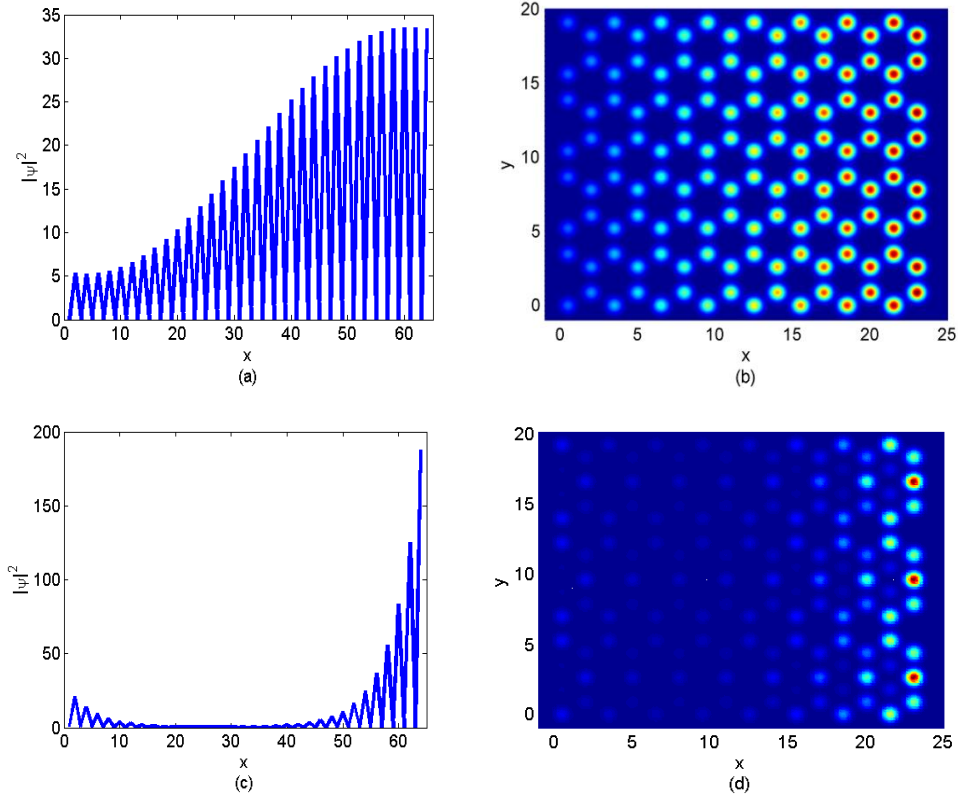

Fig.8 The local states in an armchair BN nanoribbon supercell with the line defect ($t_1 = -1.0$ eV). The size of the supercell is set as N=16, M=23.(a) The wavefunction of the local state with $p$=8 in the 1D-supercell model; (b) The electron density distribution of the local state with $p$=8 in the 2D-TB model; (c) The wavefunction of the local state with $p$=9 in the 1D-supercell model;(d): The electron density distribution of the local state with $p$=9 in the 2D-TB model .

In Figs. 8(a) and 8(b), it seems that the wavefunction decays only from the right to the left. In fact, it also decays from the left side to the right side as seen apparently in Figs.8(c) and 8(d). The decaying trend from the left side is hidden by another trend which decays from the right side. In this supercell the nitrogen atoms are on the right boundary and the boron atoms are on the left boundary. As $\varepsilon_N < \varepsilon_B$ (Table 1), we find that for the band besides the Fermi level, the electron density near the nitrogen-atom boundary (right) is larger since these atoms have a lower energy. The electron densities from both boundaries all decay towards the inner region of the supercell.

## 4. Conclusions

We have developed a type of 1D band calculation method for the GNR supercell systems, including the supercells with line defects and uniaxial strains. We have calculated the energy bands in these supercell systems. Our results are identical with the 2D-TB method except in the case of odd N number (the length of the supercell). In this case some extra bands need to be deduced. We have investigated some special decaying (local) states in the armchair graphene-like supercells and in the semi-infinite nanoribbons in this 1D method combined with the transfer matrix technique. We have derived a condition for existence of these local state in the semi-infinite aGNR: $p$>(N+1)/3. We have found that this condition can also be a criteria for the local state in the aGNR supercells with a strong line defect (a small hopping integral near the defect).

We also have studied the band structures and local states in the BN nanoribbon supercell. We have seen that there exists some asymmetric local state in BN ribbons due to the different boundary atoms in the BN supercell.

## Acknowledgement

This work is supported by the starting foundation for the 'Hundred Talent Program' of Chongqing University, China. The computer help from Dr. Rui Wang in Chongqing University is also kindly appreciated.